\documentclass[12pt]{article}

\usepackage[margin=2.5cm]{geometry}
\usepackage{graphicx}
\usepackage{amsmath,amssymb,amsfonts}
\usepackage{booktabs}
\usepackage{mathtools}
\usepackage{xcolor}
\usepackage{multirow}
\usepackage{array}
\usepackage{float}
\usepackage{cite}
\usepackage{url}

\DeclareUnicodeCharacter{2212}{-}

\title{High average flux positron sources driven by a mJ-class laser}

\begin{document}

\maketitle

% Affiliation macros
\newcommand{\QUB}{School of Mathematics and Physics, Queen's University of Belfast, BT7 1NN, Belfast, UK}
\newcommand{\ELIALPS}{ELI-ALPS, ELI-HU Non-Profit Ltd., Wolfgang Sandner utca 3, Szeged 6728, Hungary}
\newcommand{\ELIBL}{ELI Beamlines Facility, The Extreme Light Infrastructure ERIC, Dolni Brezany 25241, Czech Republic}
\newcommand{\Dundee}{Carnegie Laboratory of Physics, SUPA, School of Engineering, Physics, and Mathematics, University of Dundee, Dundee DD1 4HN, United Kingdom}

\author{
\parbox{0.9\textwidth}{\centering
T.~Foster$^{1}$,
D.~Papp$^{2}$,
C.~McAnespie$^{1}$,
H.~Maguire$^{1}$,
J.~Csontos$^{2}$,
E.~Gerstmayr$^{1}$,
G.~M.~Grittani$^{3}$,
C.~Kamperidis$^{2}$,
D.~J.~Keeble$^{4}$,
C.~M.~Lazzarini$^{3}$,
S.~Lorenz$^{3}$,
T.~Lucza$^{2}$,
S.~Tóth$^{2}$,
and G.~Sarri$^{1}$\\[1em]
$^{1}$ \QUB\\
$^{2}$ \ELIALPS\\
$^{3}$ \ELIBL\\
$^{4}$ \Dundee
}
}

\begin{abstract}
We report on the production and characterisation of MeV-scale positron populations driven by a high-repetition-rate, mJ-class laser system. The positrons are generated via a two-step Bethe--Heitler process driven by laser-wakefield-accelerated electron beams propagating through a mm-scale lead converter. At 10~Hz, approximately $10^{3}$ positrons in the 0.8--1.7~MeV energy range were measured over 100 accumulated shots. Stable electron acceleration was demonstrated at 10~Hz, 100~Hz, and 1~kHz using approximately 9~mJ of laser energy in the focal spot, indicating the potential for substantially higher average positron fluxes for laser systems operating at kHz repetition rates and energies of 10s of mJ. Such a positron source allows generating sufficiently high average fluxes of~MeV-scale positrons while maintaining low positron yields per pulse and narrow temporal distributions, a regime of particular interest for novel applications in material science and fundamental physics. The measured positron yields and spectra are in good agreement with Monte Carlo simulations, supporting the modelling of the positron-production process and magnetic transport system.

\end{abstract}

\newpage

\section{Introduction}

Laser–plasma accelerators have emerged as compact and highly tuneable sources of relativistic particle beams capable of producing electrons, ions, and photons with energies ranging from the~MeV up to the multi-GeV scale. In particular, laser-wakefield acceleration (LWFA)~\cite{Esarey2009} enables the generation of high-energy femtosecond-scale electron beams within cm-scale plasmas driven by ultrashort, high-intensity laser pulses. These electron beams have been widely explored as drivers for secondary radiation sources, including X-rays, $\gamma$-rays, and particle beams~\cite{Streeter2024,TaPhuoc2012,Corde2013,Calvin2026,Shalloo2020, Sarri2014,Mirzaie2024,Wood2026}.

Recent advances in ultrafast laser technology have enabled the development
of high-repetition-rate, TW-scale laser systems capable of operating at
hundreds of~Hz to~kHz repetition rates~\cite{Budriunas2017,Golinelli2019}.
These systems enable laser--plasma particle sources to operate at substantially higher average flux while retaining the ultrashort temporal characteristics of laser-driven acceleration~\cite{Gustas2018,Rovige2020,
Lazzarini2024}. In particular, high-repetition-rate LWFA can provide femtosecond-scale electron beams for applications including ultrafast imaging~\cite{He2013,He2016} and the generation of secondary
sources~\cite{Dpp2016,Sadighi2026,Feng2023,Feng2025,Papp2022}.

One emerging and particularly promising application of LWFA-driven electron beams is the generation of positrons through bremsstrahlung-mediated pair production. In this process, relativistic electrons interact with a high-$Z$ converter target, producing bremsstrahlung photons which subsequently generate electron–positron pairs via the Bethe–Heitler process~\cite{heitler1984}. Laser-driven positron sources have attracted significant interest for their potential applications in high-energy physics~\cite{Streeter2024,Alejo2019}, laboratory astrophysics\cite{Sarri2015,Warwick2017,Chen2015} %\textcolor{red}{cite Sarri Nat. Comm. 2015, Warwick PRL 2027, and H. Chen PRL 2015}
and materials science~\cite{Audet2021}. %\textcolor{red}{cite Audet PRAB 2021}.
Justified by these recent developments, plasma-based positron sources have been included as a flagship application of the ESFRI user facility EuPRAXIA~\cite{Assmann2020,Sarri2022}.

In materials science, a particularly important application is Positron
Annihilation Lifetime Spectroscopy (PALS), a non-invasive technique capable of
identifying sub-nanometre defects in materials~\cite{Audet2021}. An ideal
positron source for PALS should combine a high average flux, allowing
sufficient statistics to be accumulated within practical acquisition times,
with a low positron yield per pulse to minimise detector pile-up. Short positron bunch durations are also desirable in order to minimise the beam-related contribution to the instrument response function and maximise
the temporal resolution of the measurement.
Furthermore, a tuneable positron energy is required for depth-resolved studies,
with increasing positron kinetic energy enabling probing at greater depths within a
material. Current high-intensity variable-energy PALS facilities typically deliver
monoenergetic positrons with energies from approximately 0.1 to 20~keV,
enabling depth-resolved measurements only within microns of the surface of the sample. The width of their instrument response function %(IRF), acronym not needed, since it is not used elsewhere
is typically of
the order of 180--250~ps, with contributions from both the temporal spread
of the positron pulse and the detector response~\cite{Egger2010,Wagner2017}.
Laser-driven sources offer the prospect of extending precision PALS
measurements into the MeV energy range, enabling substantially greater
implantation depths. Their short intrinsic bunch duration, combined with
appropriate energy selection, could also reduce the beam-related
contribution to the timing response, while high-repetition-rate operation
offers the potential for rapid accumulation of PALS spectra.
%\textcolor{red}{Here we need a paragraph explaining that PALS is one of these applications, describe the requirements for this application (high average flux but low yield per pulse, short duration, and tuneability in energy for volumetric studies) and discuss how conventional sources are limited in these respects. You can use the introduction of Audet PRAB2021}.

%particularly in Positron Annihilation Lifetime Spectroscopy (PALS)~\cite{PULSTAR,NEPOMUC,PENELOPE}, where short positron bursts with tunable energies could enable time-resolved studies of nanometer defects in solids.

The first demonstrations of laser-driven, MeV-scale positron bunches were
achieved using high-energy, single-shot laser pulses focused directly onto
suitable high-$Z$ solid targets~\cite{Chen2010,Audet2021}. In this scheme,
the laser generates super-thermal electron populations within the target,
which subsequently drive positron generation while propagating through the
remainder of the converter. These experiments generally require high laser
pulse energies and therefore operate at low repetition rates, limiting the
achievable average positron flux and typically requiring large-scale
high-energy laser facilities. Audet et al.~\cite{Audet2021} numerically proposed an
alternative approach in which positron generation is driven by
laser-wakefield-accelerated electrons produced using mJ-class,
high-repetition-rate laser systems. Provided that the converter can sustain
high-repetition-rate operation, this approach offers a route towards more
compact sources with high average positron flux.

In this work, we report on the first experimental demonstration of laser-driven positron production using a table-top~mJ-class laser system. The experiment was conducted at the Extreme Light Infrastructure--Attosecond Light Pulse Source (ELI-ALPS) using the eSYLOS beamline, driven by the
SYLOS2 laser system, delivering 25~mJ, 14~fs pulses with a repetition rate tuneable from 10~Hz to 1~kHz~\cite{eliSylos3}. Approximately 9~mJ of laser energy was contained within the focal spot on
a 300 micron gas jet, generating multi-MeV electron beams that subsequently propagated through a mm-scale lead converter. The generated positrons were captured by a quadrupole doublet and magnetically dispersed onto a scintillator screen. Operations at 10~Hz resulted in $>10^3$ positrons accumulated over a 100 shots, with energies ranging from 0.8 to 1.7~MeV. While converter ablation precluded positron measurements at higher repetition rates, stable electron beams were demonstrated at 100~Hz and 1~kHz. Assuming a comparable positron yield per shot, operation at 1~kHz would correspond to an average detected positron flux exceeding $10^4$ positrons per second, with higher positron yields achievable with laser energies approaching 100 mJ. The experimental measurements are in quantitative agreement with FLUKA~\cite{FLUKA} Monte Carlo simulations, supporting the modelling of the positron-production process and magnetic transport system. These results demonstrate the possibility of producing high flux and tuneable MeV-scale positrons using a table-top high repetition rate laser, with characteristics of relevance to material science applications. 

%\newpage
\section{Experimental Setup}

\begin{figure}[t!]
\centering
\includegraphics[width=1\linewidth]{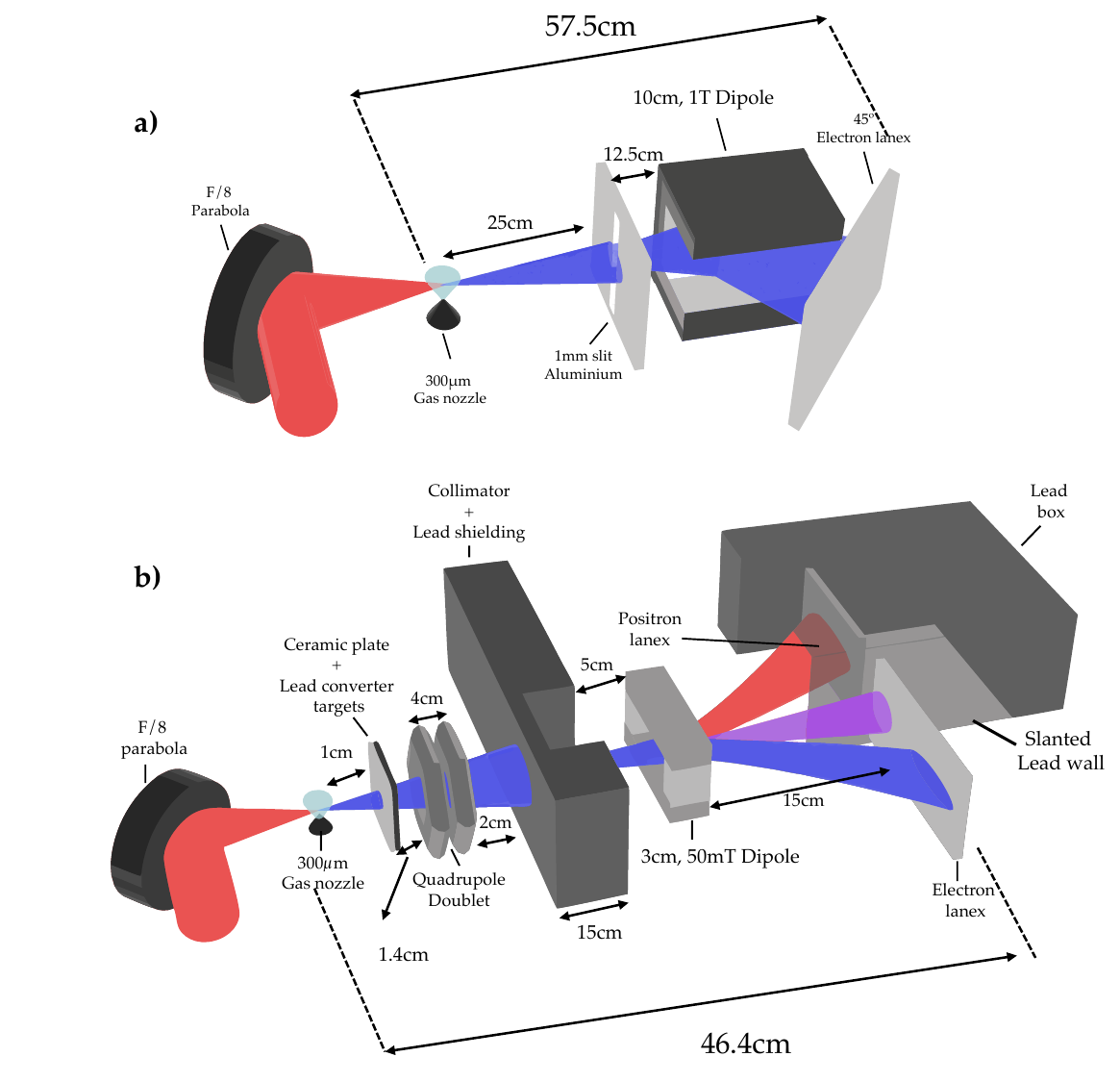}

\caption{Schematics of experimental setup for a) High repetition wakefield characterisation and b) Positron generation.}
\label{Fig1}
\end{figure}

Two experimental configurations were used sequentially. The first,
shown in Fig.~\ref{Fig1}a), was used to characterise the LWFA electron
beams driving the positron-production process. The second,
shown in Fig.~\ref{Fig1}b), incorporated the converter, quadrupole
doublet, shielding, and positron spectrometer for the positron
measurements. In both configurations, the laser delivered 36~mJ before compression, corresponding to approximately 25~mJ after compression. %for a compressor throughput of approximately 70\%. 
The 14~fs pulses, at a central wavelength
of 836~nm, were focused by an $f/8$ off-axis parabolic mirror
onto a gas target to drive laser-wakefield acceleration. The pre-compression laser energy varied by approximately $\pm2$~mJ
($\pm6\%$) about its mean value over the course of the experimental run. The measured focal-spot dimensions were
$(6.89\pm0.10)\times(7.10\pm0.09)~\mu$m at 10~Hz,
$(7.04\pm0.13)\times(7.57\pm0.16)~\mu$m at 100~Hz, and
$(7.03\pm0.10)\times(7.16\pm0.21)~\mu$m at 1~kHz,
where the quoted uncertainties are the $1\sigma$ uncertainties obtained
from the Gaussian fits. The corresponding energies contained within the
focal spot were 9.5, 9.0, and 9.2~mJ, respectively. Assuming Gaussian spatial and temporal profiles and a pulse duration of
14~fs, these parameters correspond to a mean estimated peak intensity of
$(2.15\pm0.15)\times10^{18}~\mathrm{W\,cm^{-2}}$ across the three
operating conditions, where the quoted spread is the standard deviation between the three configurations. The corresponding normalised vector
potential $a_0$ is of the order of 1. % This bit is asking for trouble... The focal-spot diagnostic was subsequently found to be affected by a damaged neutral-density filter; the small differences between the measured profiles at different repetition rates should therefore not be interpreted as direct measurements of changes in the on-target focal spot.
The laser was focused onto a supersonic gas jet with a diameter of $300~\mu$m, delivering pure nitrogen at a plasma density of $10^{19}$~cm$^{-3}$. 

In the first part of the experiment, the electron beams were characterised using a magnetic spectrometer comprising a 1~mm-wide slit in a 1~mm-thick aluminium plate, a dipole magnet, and a Scintacor Luminex UltraBright~\cite{ScintacorLuminex} screen. Two dipole magnets were used: a 1~T, 10~cm dipole to resolve electron energies above 6.5~MeV and a 50~mT, 3~cm dipole to measure electron energies from 0.8 to 9~MeV.

For the positron generation (Fig. \ref{Fig1}.b) a converter target comprising a 1~mm ceramic plate and a lead sheet of thicknesses ranging from 0.5 to 2~mm was positioned 1~cm downstream of the gas-jet. The ceramic plate was used to mitigate direct laser ablation of the lead target and was placed at $\simeq15^\circ$ to avoid laser back-reflection into the laser chain. Two permanent-magnet quadrupoles, with aperture diameters of 6.1 and 6.2~cm
and corresponding field gradients of 26.23 and 12.9~T/m, respectively, were
placed downstream of the converter and separated by 4~cm to partially
collimate the positron beam. The positrons were subsequently spatially separated and spectrally resolved using the 50~mT dipole magnet. For the converter measurements, the electron-side spectrometer covered an energy range of 1.26--6.5~MeV. The reconstructed spectra were sampled with an energy-bin spacing of approximately 0.23~MeV, while the calculated spectrometer energy resolution, $\Delta E$, varied
from approximately 8~keV at 1.26~MeV to 224~keV at 6.5~MeV. Two separate scintillator screens were used to simultaneously measure the positrons and the scattered electrons from the converter. The scintillator screens were imaged by Andor iXon Ultra 888 EMCCD cameras. Each image was acquired using a 10s exposure time at a laser repetition rate of 10~Hz, corresponding to an integration over 100 laser shots. Dark-frame and background subtraction were applied to the recorded images.

A shielding configuration and collimator comprising 50-mm-thick lead bricks was used to
minimise background arising from scattered electrons and bremsstrahlung
photons in the positron detection area.

\newpage
\section{Electron Beam Characterisation}

\begin{figure}[t!]
\centering
\includegraphics[width=0.9\linewidth]{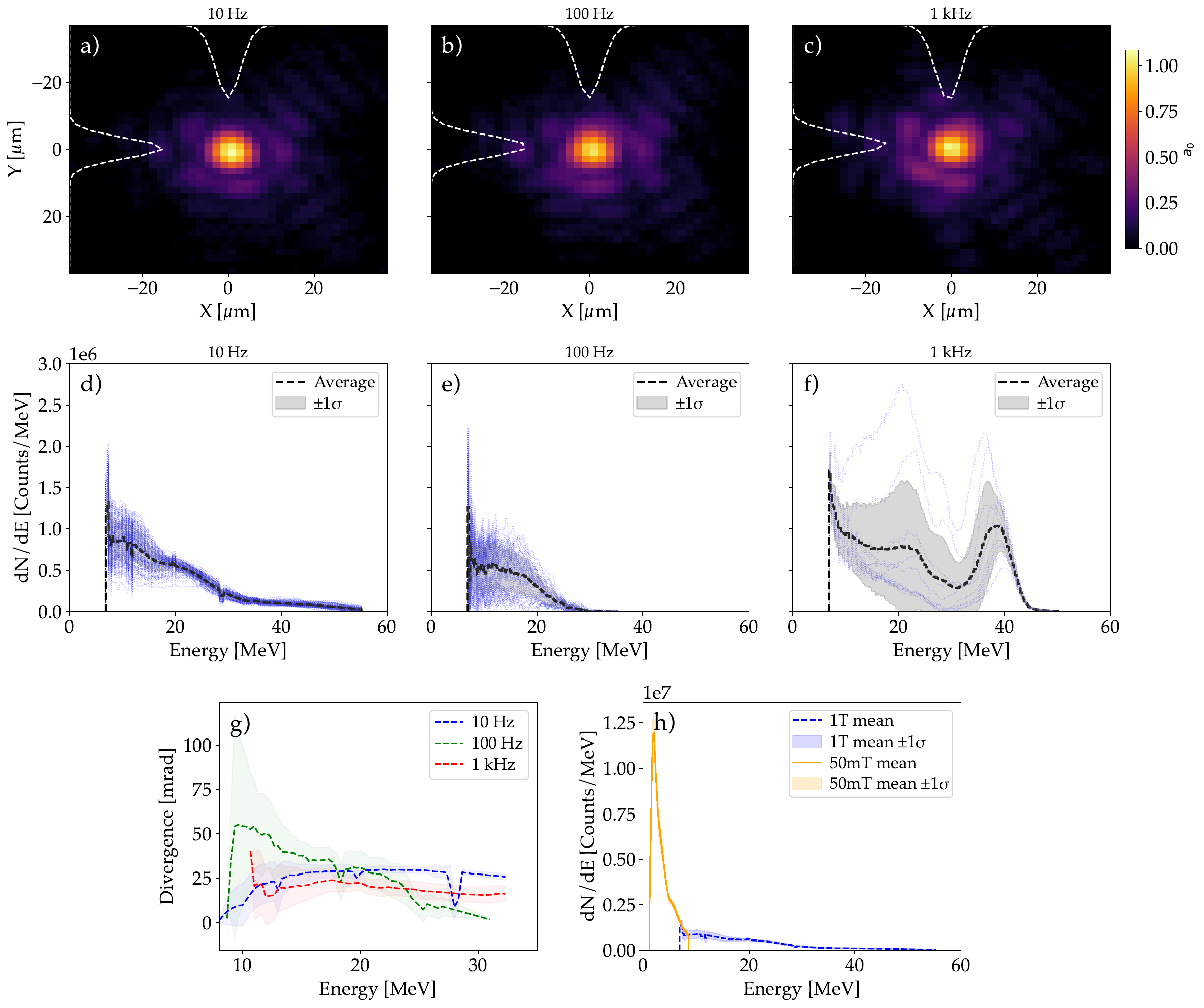}
\caption{
\textbf{a)--c)} Spatial distribution of the laser normalised vector
potential in the focal spot measured at 10~Hz, 100~Hz, and 1~kHz,
respectively.
\textbf{d)--f)} 50 individual single-shot electron spectra measured at
each repetition rate. Blue lines indicate the individual spectra, while
the dashed black line indicates their average and the grey shading
represents the standard deviation.
\textbf{g)} Mean electron-beam divergence as a function of energy. Dashed lines indicate the mean divergence over 50 consecutive shots and shaded
regions represent $\pm1\sigma$.
\textbf{h)} Full electron spectrum at 10~Hz obtained by combining
measurements with both dipole magnets.
}
\label{Fig2}
\end{figure}

The LWFA electron beam was characterised at three laser repetition rates
using the 1~T dipole magnet. Figs.~\ref{Fig2}d--f show the single-shot
electron spectra for each operating condition, together with their average
and standard deviation. All three operating conditions produced reproducible broad electron spectra
extending to approximately 30--50~MeV. The main qualitative difference was
the appearance of a higher-energy feature around 40~MeV during 1~kHz
operation. This feature is attributed to the additional optimisation of the
laser focal spot that was carried out before operations at 1~kHz. FLUKA simulations indicate
that this feature has a negligible effect on the characteristics of the
generated positron population. % not needed... Variations in the actual on-target focal spot may therefore have contributed to the differencesobserved in the electron spectra. 
% this is a repetition... FLUKA simulations indicate that the 40~MeV feature has a negligible effect on the characteristics of the generated positrons. 
The measured electron spectra are consistent with the numerical predictions reported in Ref.~\cite{Audet2021} for LWFA driven by a~mJ-class,~kHz laser system at ELI-ALPS. Their particle-in-cell simulations
predict a broad MeV-scale electron distribution, with a substantial
low-energy population and a high-energy tail extending to approximately
60~MeV, comparable to the broad distributions extending to 30--50~MeV
observed experimentally here.
Electron beams with a comparable energy scale have also recently been demonstrated experimentally on the same beamline using the higher-power
SYLOS3 laser system, with maximum electron energies of approximately 47~MeV
reported during stable 100~Hz operation~\cite{Smartsev2026}.

The energy-resolved divergence of the electron beam along the non-dispersed
axis of the spectrometer is also stable across the different repetition rates
(Fig.~\ref{Fig2}g), with values of the order of 20--30~mrad and only a weak
dependence on electron energy. % do we need this?....The shot-to-shot electron-beam pointing fluctuations, defined as the standard deviation of the beam centroid position over 10 shots, were $(\sigma_x,\sigma_y)=(1.57,5.13)$~mrad at 10~Hz, $(0.52,1.22)$~mrad at 100~Hz, and $(0.63,6.24)$~mrad at 1~kHz.

To extract information on the electron spectra for energies below 6.5 MeV, a series of spectra
were acquired at 10~Hz using the 50~mT, 3~cm dipole magnet, which provided
an energy range of 0.8--9~MeV. These measurements indicated a large
population of electrons at MeV-scale energies. The spectra obtained using
the 50~mT and 1~T dipole magnets overlap between 6.5 and 9~MeV, allowing
the two measurements to be combined to reconstruct the full electron
spectrum shown in Fig.~\ref{Fig2}h, which was subsequently used as input
to the Monte-Carlo FLUKA simulations. Charge calibration of the magnetic
spectrometer using a tritium source indicated that the electron beam
typically contained approximately 8~pC per shot across the full spectrum, with
approximately 5~pC contained below 5~MeV.

\section{Positron production and characterisation}

Following characterisation of the LWFA electron beam, a converter target comprising a lead foil and a 1~mm ceramic plate was inserted and the setup modified (Fig. \ref{Fig1}.b). Lead converter thicknesses of 0.5, 1, 1.5, and 2~mm were investigated experimentally.

\begin{figure}[H]
\centering
\includegraphics[width=1\linewidth]{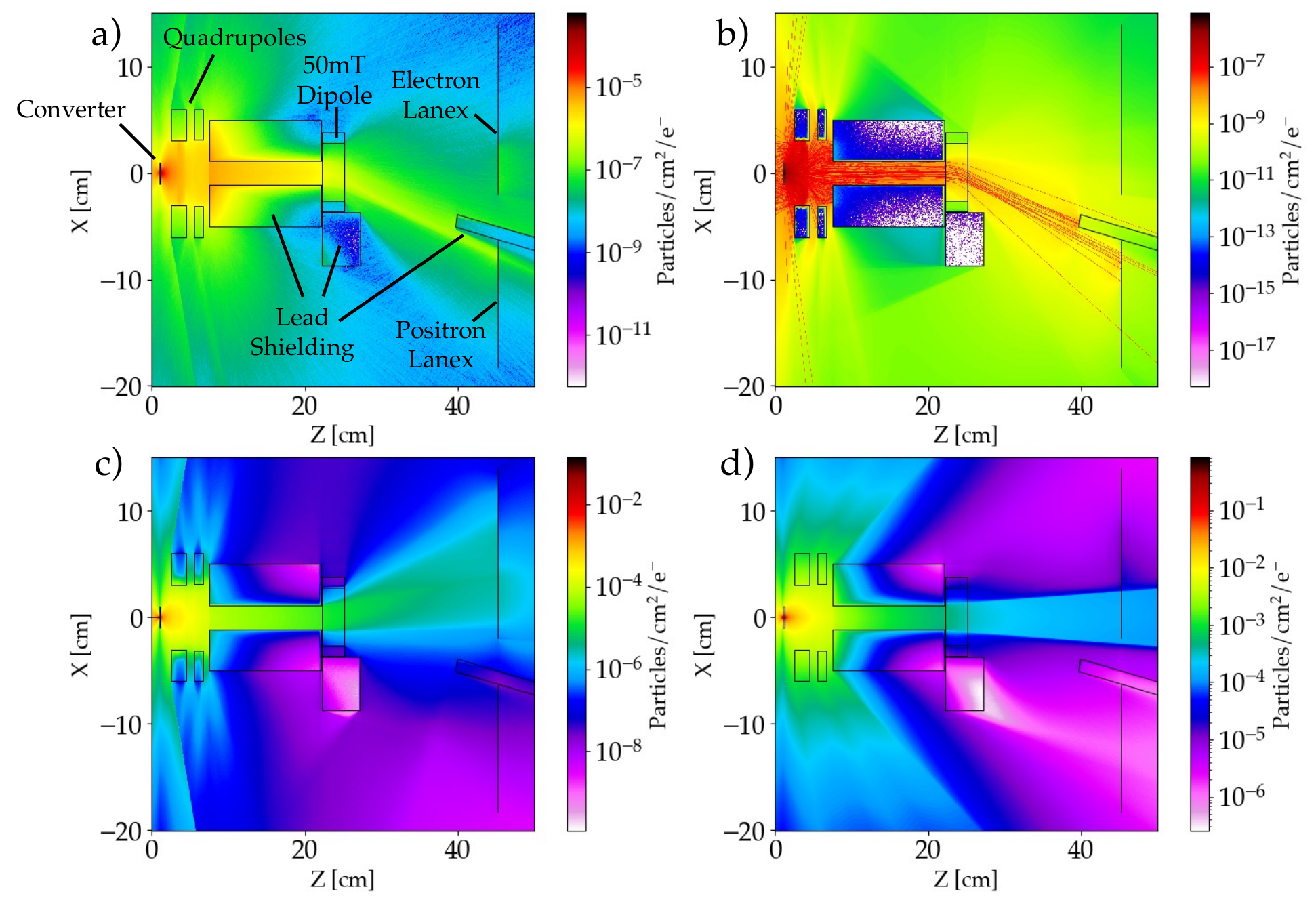}
\caption{Side-view simulated particle distributions for a 2~mm Pb converter of
\textbf{a)} positrons with kinetic energies above 100~keV,
\textbf{b)} positrons with kinetic energies between 0.5 and 2~MeV,
\textbf{c)} electrons with kinetic energies above 100~keV, and
\textbf{d)} photons with energies above 10~keV.
The distributions are scored over a 10~cm transverse extent
($-5<y<5$~cm). Colorbars are expressed in particles per cm$^2$ per primary
electron.}
\label{Fig3}
\end{figure}

Monte-Carlo simulations using the FLUKA particle transport code
\cite{FLUKA} were first performed to estimate the expected positron yield,
spectrum, and signal-to-noise ratio in the positron detection area. The simulations used the FLUKA \texttt{EM-CASCA} defaults, corresponding to transport thresholds of 100~keV in kinetic energy for electrons and positrons and 10~keV for photons. As input
to the simulations, the experimentally measured electron spectrum shown in
Fig.~\ref{Fig2}h was used together with a FWHM divergence of 30~mrad, representative
of the experimentally measured electron-beam divergence shown in
Fig.~\ref{Fig2}g. Each simulation comprised $8\times10^{6}$ primary
electrons, with the reported results accumulated over 1000 independent runs,
corresponding to a total of $8\times10^{9}$ simulated primary electron
histories. The transverse distributions of generated positrons, scattered
electrons, and photons are shown in Fig.~\ref{Fig3} for a 2~mm lead
converter, with the main elements of the setup labelled in panel a).
Simulations were also performed for each of the experimentally investigated
converter thicknesses. For the thinner converters, the simulations indicated an insufficient
signal-to-noise ratio for reliable positron detection, as also confirmed
experimentally.

The simulations confirm efficient capture and collimation of positrons with energies around 1~MeV, with approximately $2\times10^{-7}$ positrons per primary electron reaching the scintillator screen (see Figs. \ref{Fig3}.a and \ref{Fig3}.b). For an initial electron beam charge of 8 pC, this corresponds to approximately 10 positrons reaching the detector per shot. The shielding configuration is effective at shielding the positron detection area from the bremsstrahlung photons and the scattered electrons.

\begin{figure}[t!]
\centering
\includegraphics[width=1\linewidth]{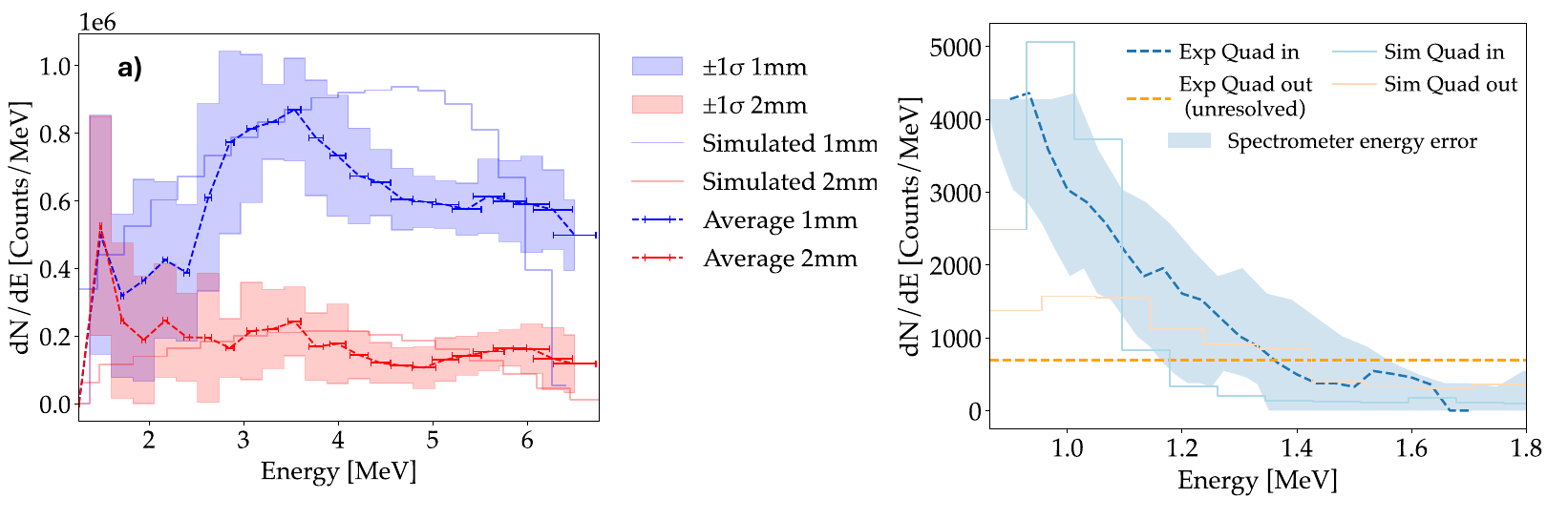}
\caption{\textbf{a)} Measured spectra of scattered electrons after the converter
for 1~mm (blue) and 2~mm (red) converter thicknesses, averaged over 10 shots.
The shaded regions indicate the corresponding standard deviation, while the
horizontal error bars represent the energy uncertainty. The corresponding
FLUKA-simulated spectra are shown as thin solid lines for comparison.
\textbf{b)} Measured positron spectra using a
2~mm lead converter with and without the quadrupole doublet. With the
quadrupoles inserted, the shaded region indicates the calculated
energy-dependent resolution of the magnetic spectrometer. Without the
quadrupoles, the increased angular acceptance results in an energy resolution
larger than the measurable energy range, preventing reconstruction of an
energy-resolved spectrum; the measured signal is therefore represented by
a constant value across the detectable energy interval. The corresponding
FLUKA spectra with (blue) and without (orange) the quadrupole magnets are overplotted for
comparison.}
\label{Fig4}
\end{figure}

The measured energy spectra of the scattered electrons and the generated
positrons are shown in Fig.~\ref{Fig4} for a 10-second acquisition time and
10~Hz operation. Positron production at higher repetition rates was
limited by converter ablation and degradation, which resulted in
contamination of the focusing parabola. Strategies to contain this issue and operate at higher repetition rates are discussed in the Conclusions section.

% Extending positron production to
% higher repetition rates will therefore require mitigation of target
% degradation, for example using replenishable or flowing target approaches
% such as tape-drive or liquid-sheet systems~\cite{Biagioni2025EuPRAXIA,
% Xu2023,Morrison2018}.

The spectrum of the scattered electrons could be detected above noise for 1mm and 2mm converter thicknesses. As intuitively expected, the scattered electrons have a relatively flat spectrum, with a higher number of electrons detected for the thinner converter target. The spectra are in good agreement with FLUKA simulations, validating the numerical modelling of the experiment. 

Positron spectra could only be measured above noise using a 2-mm-thick converter target. Increasing the converter thickness enhances the pair-production probability but also increases bremsstrahlung emission, multiple scattering, and positron
attenuation within the target. The resulting detected signal therefore depends on the balance between increased positron production and increased background and transport losses. Within the 0.5--2~mm range investigated
experimentally, the 2~mm converter provided the strongest detectable positron signal and was therefore selected for detailed positron characterisation. The positron spectrum is measured above noise between 0.8 and 1.7 MeV, where the highest number of positrons is expected to be transported by the specific configuration of quadrupoles and dipole. Within this energy range, the positron spectrum has a monotonically decreasing shape, in good agreement with FLUKA simulations including the quadrupole magnets. The inclusion of the quadrupole doublet significantly enhances the detected
positron signal around 1~MeV by improving the capture of low-energy positrons
emerging from the converter. Removing the quadrupoles decreases the detected
positron flux by a factor of approximately 3.5. Without the quadrupoles, the larger angular distribution of the positron population results in an energy
resolution larger than the measurable energy range, preventing reconstruction
of an energy-resolved spectrum. The measurement is therefore represented by
a constant value across the detectable energy interval, with only the
integrated positron yield retained.

To estimate the number of positrons measured over a 10-second acquisition time
(accumulation over 100 shots during operation at a 10~Hz repetition rate),
the scintillator response was calibrated following the procedure outlined
by Glinec et al.~\cite{Glinec2006}. The scintillator screens used in
the experiment employ a Gd$_2$O$_2$S:Tb phosphor with a peak emission
wavelength of 545~nm and a manufacturer-specified intrinsic efficiency of
15\%~\cite{ScintacorLuminex}. This is consistent with the scintillation
efficiency of $(16\pm2)\%$ reported by Glinec et al.~\cite{Glinec2006}
and adopted here for the calibration. The main parameters used in the calibration are summarised in
Table~\ref{tab:lanex_calibration_values}, resulting in a calibration factor
of $(130\pm20)$ counts/positron on the camera. Following this calibration,
the experimental spectrum shown in Fig.~\ref{Fig4}b corresponds to an
integrated positron yield of approximately $(1.0\pm0.2)\times10^{3}$ positrons in the 0.8--1.7~MeV energy range accumulated over 100 shots, in good quantitative
agreement with FLUKA simulations.

The positron bunch duration at the converter rear surface is expected to be
comparable to that of the primary electron beam ($\sim50$~fs)
\cite{Audet2021}. During propagation, however, the broad positron
energy distribution produces an additional time-of-flight spread. Over a
propagation distance of 46.4~cm, accepting the full measured kinetic-energy
range of 0.8--1.7~MeV corresponds to a time-of-flight spread of approximately
90~ps. This value represents the temporal spread associated with the full
accepted energy interval rather than the intrinsic positron bunch duration.
Further energy selection would reduce this contribution and could therefore
provide arrival-time significantly below 100~ps at the sample. The resolution of a PALS instrument utilising such a positron source would thus be ultimately limited by the detector response only.

%\section{Positron Characterisation}
\begin{table}[t!]
\centering
\small
\begin{tabular}{lll}
\hline
\textbf{Parameter} & \textbf{Symbol} & \textbf{Value Used} \\
\hline
CCD quantum efficiency at 545\,nm
& QE
& $0.95 \pm 0.02$ \\

CCD conversion factor & $r$ & 16.1 electrons/count \\
Optical transmission (window $\times$ lens)
& $T$
& $0.87 \pm 0.02$ \\

Collection solid angle
& $\Omega$
& $(1.91 \pm 0.06)\times10^{-3}$ sr \\

Scintillation efficiency
& $\epsilon$
& $0.16 \pm 0.02$~\cite{Glinec2006,ScintacorLuminex} \\

Energy of scintillation photon
& $E_\gamma$
& 2.27 eV \\

Gd$_2$O$_2$S active medium density
& $\rho_{\mathrm{GOS}}$
& $7.44\,\mathrm{g\,cm^{-3}}$~\cite{Glinec2006} \\

Assumed phosphor surface loading
& $h_s$
& $33\,\mathrm{mg\,cm^{-2}}$~\cite{Glinec2006} \\

Assumed effective phosphor thickness
& $\delta_x$
& $44.4\,\mu\mathrm{m}$~\cite{Glinec2006} \\

Energy loss in phosphor (for 1~MeV)
& $dE/dx$
& $8.9\,\mathrm{MeV\,cm^{-1}}$~\cite{estar} \\
\hline
Resulting calibration factor
& --
& $(130 \pm 20)$ counts/positron \\
\hline
\end{tabular}
\caption{Main parameters used for the absolute positron-yield calibration of
the Scintacor Luminex UltraBright scintillator screen and imaging system,
following the procedure of Glinec et al.~\cite{Glinec2006}. The quoted
uncertainty in the calibration factor was obtained by propagating the
uncertainties associated with the scintillation efficiency, optical
transmission, collection solid angle, and detector response.}
\label{tab:lanex_calibration_values}
\end{table}

\section{Conclusion}

We report on the first experimental demonstration of positron production at high repetition rate using a table-top setup driven by a~mJ-class laser system. The positrons are generated following a two-step Bethe-Heitler process driven by laser-wakefield accelerated electron beams propagating through a mm-scale lead converter. 

After transport, measured positron yields above noise of approximately $10^{3}$ positrons in a 0.8--1.7~MeV energy range were obtained after the accumulation of 100 laser shots at a 10~Hz repetition rate. The measured yield and spectral distribution of the positrons is in good agreement with matching Monte-Calro simulations, validating the modelling of the converter, magnetic transport system, and detector response.

The demonstrated stable operation of the LWFA electron source up to 1~kHz
indicates the potential for achieving average positron fluxes exceeding
$10^4$ positrons per second, provided that the positron-production stage
can be extended to kHz operation while maintaining a comparable yield per
shot. Higher laser energies, reaching approximately 60~mJ in focus with
the SYLOS3 laser at ELI-ALPS, provide scope for further optimisation of the
LWFA electron beam. However, the resulting positron yield cannot be inferred
from laser-energy scaling alone, since it depends on the electron-beam
charge, spectrum, and divergence.

In the present experiment, positron operation above 10~Hz was limited by
converter degradation and debris contamination of the focusing optic.
Extending the positron-production stage to higher repetition rates will
therefore require both improved target management and debris mitigation.
Replenishable target concepts, such as tape-drive or flowing-target systems,
could mitigate progressive target degradation~\cite{Biagioni2025EuPRAXIA,Xu2023,Morrison2018}, while appropriate debris-mitigation measures would be
required to protect the focusing optic. Such developments would allow
positron production to be tested at the 100~Hz--kHz repetition rates already
demonstrated for the LWFA electron source.

The positron beam properties are of relevance for the volumetric
characterisation of materials using positron annihilation lifetime
spectroscopy. The broad positron spectrum at source allows further energy
selection to provide the kinetic energy required for a given application.
Narrowing the accepted energy interval would also reduce the beam-related
time-of-flight spread. The achievable temporal resolution is therefore
dependent on the selected positron energy bandwidth, propagation distance,
and detector timing response, with the latter ultimately setting the
instrumental limit once the beam-related contribution becomes sufficiently
small. Combined with the MeV-scale positron energies demonstrated here, this could enable
PALS measurements at substantially greater implantation depths than
conventional keV-scale variable-energy positron beams, complementing
existing facilities designed primarily for near-surface characterisation
\cite{Egger2010,Wagner2017}.

In conclusion, the results reported here demonstrate MeV-scale positron production using a
high-repetition-rate, mJ-class LWFA platform. While positron measurements in the present experiment were limited to 10~Hz by converter degradation, stable LWFA electron-beam generation was demonstrated up to 1~kHz. Extending the converter and debris-mitigation system to kHz operation is therefore the
key experimental step required to translate the demonstrated per-shot
positron yield into substantially higher average positron fluxes.\\

\textbf{Acknowledgments}: We acknowledge the Extreme Light Infrastructure
(ELI) for providing access to experimental facilities and support. The
experiments were performed at the eSYLOS beamline at the ELI-ALPS Facility,
with beamtime allocated under the 4th ELI ERIC User Call
(Experiment ELIUPM4-41). The use of instrumentation, services (including
sample preparation), and experimental support was made possible by the ELI
Excellence-based User Programme. G. Sarri wishes to acknowledge support from
the EPSRC (grant No. EP/V044397/1) and funding from the EUPRAXIA-PP
HORIZON-INFRA-2021-DEV-02 EUR project 101079773.

% Bibliography
\bibliographystyle{unsrt}
\bibliography{Refs}

\end{document}